\documentclass[10pt,twocolumn,prl,aps,amssymb,amsmath,showpacs,tightenlines,floatfix]{revtex4}
\usepackage{psfig}

\newcommand{\re}{\mbox{$\rm e$}}
\newcommand{\ri}{\mbox{$\rm i$}}
\newcommand{\rd}{\mbox{$\rm d$}}

\begin{document}

\title{Quantum Phase Transitions Without Thermodynamic Limits}

\author{Dorje~C.~Brody${}^*$, Daniel~W.~Hook${}^*$, and
Lane~P.~Hughston${}^\dagger$}

\affiliation{${}^*$Blackett Laboratory, Imperial College, London
SW7 2BZ, UK \\ ${}^\dagger$Department of Mathematics, King's
College London, The Strand, London WC2R 2LS, UK}

\date{\today}

\begin{abstract}
A new microcanonical equilibrium state is introduced for quantum
systems with finite-dimensional state spaces. Equilibrium is
characterised by a uniform distribution on a level surface of the
expectation value of the Hamiltonian. The distinguishing feature
of the proposed equilibrium state is that the corresponding
density of states is a continuous function of the energy, and
hence thermodynamic functions are well defined for finite quantum
systems. The density of states, however, is not in general an
analytic function. It is demonstrated that generic quantum systems
therefore exhibit second-order (continuous) phase transitions at
finite temperatures.
\end{abstract}

\pacs{05.30.-d, 05.30.Ch, 45.20.Jj}

\maketitle

The derivation of phase transitions in quantum statistical
mechanics typically requires the introduction of a thermodynamic
limit, in which the number of degrees of freedom of the system
approaches infinity. This limit is needed because the free energy
of a finite system is analytic in the temperature. But phase
transitions are associated with the breakdown of the analyticity
of thermodynamic functions such as the free energy. Hence in the
canonical framework the thermodynamic limit is required to
generate phase transitions. Although the existence of this limit
has been shown for various systems (see, e.g., \cite{ruelle}), the
procedure can hardly be regarded as providing an adequate
description of critical phenomena.

One can consider, alternatively, a derivation based on the
microcanonical ensemble. The usual construction of this ensemble
\cite{huang} is to define the entropy by setting $S=k_B \ln n_E$,
where $n_E$ is the number of energy levels in a small interval
$[E,E+\Delta E]$. The temperature is then obtained from the
thermodynamic relation $T\rd S=\rd E$. This approach, however, is
not well formulated because (a) it relies on the introduction of
an arbitrary energy band $\Delta E$, and (b) the entropy is a
discontinuous function of the energy. To resolve these
difficulties, a scheme for taking the thermodynamic limit in the
microcanonical framework was introduced in \cite{griffiths}. For
finite systems, however, the difficulties have remained
unresolved.

The purpose of this paper is to demonstrate the following: (i) if
the microcanonical density of states is defined in terms of the
relative volume, in the space of pure quantum states, occupied by
the states associated with a given energy expectation  $E$, then
the entropy of a finite-dimensional quantum system is a continuous
function of $E$, and the temperature of the system is well
defined; and (ii) the density of states so obtained is in general
not analytic, and thus for generic quantum systems predicts the
existence of second-order phase transitions, \emph{without the
consideration of thermodynamic limits}.

It is remarkable in this connection that similar types of
second-order transitions have been observed recently for classical
spin systems, for which the associated configuration space
possesses a nontrivial topological structure~\cite{kastner}.

The paper is organised as follows. We begin with the analysis of
an idealised quantum gas to motivate the introduction of a new
microcanonical distribution. This leads to a natural definition of
the density of states $\Omega(E)$. Unlike the number of
microstates $n_E$, the microcanonical density $\Omega(E)$ is
continuous in $E$. As a consequence, we are able to determine the
energy, temperature, and specific heat of elementary quantum
systems, and work out their properties. In particular, we
demonstrate that in the case of an ideal gas of quantum particles,
each particle being described by a finite-dimensional state-space,
the system exhibits a second-order phase transition, where the
specific heat decreases abruptly.

\textbf{Ideal gas model}. Let us consider a system that consists
of a large number $N$ of identical quantum particles (for
simplicity we ignore issues associated with spin-statistics). We
write ${\hat H}_{\rm total}$ for the Hamiltonian of the composite
system, and ${\hat H}_i$ ($i=1,2,\ldots,N$) for the Hamiltonians
of the individual constituents of the system. The interactions
between the constituents are assumed to be weak, and hence to a
good approximation we have ${\hat H}_{\rm total} =\sum_{i=1}^N
{\hat H}_i$. We also assume that the constituents are
approximately independent and thus disentangled, so that the wave
function for the composite system is approximated by a product
state.

If the system as a whole is in isolation, then for equilibrium we
demand that the total energy of the composite system should be
fixed at some value $E_{\rm total}$. In other words, we have
$\langle{\hat H}_{\rm total} \rangle=E_{\rm total}$. It follows
that $\sum_{i=1}^N \langle{\hat H}_i \rangle=E_{\rm total}$. Now
consider the result of a hypothetical measurement of the energy of
one of the constituents. In equilibrium, owing to the effects of
the weak interactions, the state of each constituent should be
such that, on average, the result of an energy measurement should
be the same. That is to say, in equilibrium, the state of each
constituent should be such that the expectation value of the
energy is the same. Therefore, writing $E=N^{-1}E_{\rm total}$, we
conclude that in equilibrium the gas has the property that
$\langle{\hat H}_i\rangle=E$. That is to say, the state of each
constituent must lie on the energy surface ${\mathcal E}_E$ in the
pure-state manifold for that constituent. Since $N$ is large, this
will ensure that the uncertainty in the total energy of the
composite system, as a fraction of the expectation of the total
energy, is vanishingly small. Indeed, it follows from the
Chebyshev inequality that
\begin{eqnarray}
{\rm Prob}\left[ \frac{|{\hat H}_{\rm total}-E_{\rm total}|}
{|E_{\rm total}|}>x\right] \leq \frac{1}{Nx^2} \frac{\langle
({\hat H}_i-\langle{\hat H}_i\rangle)^2\rangle}{\langle{\hat H}_i
\rangle^2} \label{eq:2}
\end{eqnarray}
for any choice of $x>0$. Therefore, for large $N$ the energy
uncertainty of the composite system is negligible.

For convenience, we can describe the distribution of the various
constituent pure states, on their respective energy surfaces, as
if we were considering a probability measure on the energy surface
${\mathcal E}_E$ of a single constituent. In reality, we have a
large number of approximately independent constituents; but owing
to the fact that the respective state spaces are isomorphic we can
represent the behaviour of the aggregate system with the
specification of a probability distribution on the energy surface
of a single ``representative'' constituent.

\textbf{Microcanonical equilibrium}. In equilibrium, the
distribution is uniform on the energy surface, since the
equilibrium distribution should maximise an appropriate entropy
functional on the set of possible probability distributions on
${\mathcal E}_E$. From a physical point of view we can argue that
the constituents of the gas approach an equilibrium as follows: On
the one hand, weak exchanges of energy result in all the states
settling on or close to the energy surface; on the other hand, the
interactions will induce an effectively random perturbation in the
Schr\"odinger dynamics of each constituent, causing it to undergo
a Brownian motion on ${\mathcal E}_E$ that in the long run induces
uniformity in the distribution on ${\mathcal E}_E$. We conclude
that the equilibrium configuration of a quantum gas is represented
by a uniform measure on an energy surface of a representative
constituent of the gas.

The theory of the quantum microcanonical equilibrium state
presented here is analogous in many respects to the symplectic
formulation of the classical microcanonical ensemble described in
\cite{ehrenfest}. There is, however, a subtle difference.
Classically, the uncertainty in the energy is fully characterised
by the statistical distribution over the phase space, and for a
microcanonical distribution with support on a level surface of the
Hamiltonian the energy uncertainty vanishes. Quantum mechanically,
however, although the statistical contribution to the energy
variance vanishes, there remains an additional purely
quantum-mechanical contribution.  Hence, although the energy
uncertainty for the composite system is negligible for large $N$,
the energy uncertainties of the constituents will not in general
vanish. An expression for $\Delta H$ will be given in equation
(\ref{eq:13}) below.

\textbf{Density of states}.  To describe the equilibrium represented
by a uniform distribution on the energy surface ${\mathcal E}_E$, it
is convenient to use the symplectic formulation of quantum
mechanics. Let  ${\mathcal H}$ denote the Hilbert space of states
associated with a constituent. We assume that the dimension of
${\mathcal H}$ is $n+1$. The space of rays through the origin of
${\mathcal H}$ is a manifold $\Gamma$ equipped with a metric and a
symplectic structure. The expectation of the Hamiltonian along a
given ray of ${\mathcal H}$ then defines a Hamiltonian function
$H(\psi)=\langle\psi |{\hat H}_i|\psi\rangle/\langle\psi
|\psi\rangle$ on $\Gamma$, where the ray $\psi\in\Gamma$ corresponds
to the equivalence class $|\psi \rangle \sim \lambda|\psi\rangle$,
$\lambda\in{\mathbb C}\backslash 0$. The Schr\"odinger evolution on
${\mathcal H}$ is a symplectic flow on $\Gamma$, and hence we may
regard $\Gamma$ as the quantum phase space. Our approach to quantum
statistical mechanics thus unifies two independent lines of enquiry,
each of which has attracted attention in recent years: the first of
these is the ``geometric'' or ``dynamical systems'' approach to
quantum mechanics, which takes the symplectic structure of the space
of pure states as its starting point~\cite{kibble}; and the second
of these is the probabilistic approach to the foundations of quantum
statistical mechanics in which the space of probability
distributions on the space of pure states plays a primary role
\cite{jaynes}.

The level surface ${\mathcal E}_E$ in $\Gamma$ is defined by
$H(\psi)=E$. The entropy associated with the corresponding
microcanonical distribution is $S(E)=k_B\ln \Omega(E)$, where
\begin{eqnarray}
\Omega(E)=\int_{\Gamma}\delta(H(\psi)-E)\rd V_{\Gamma}.
\label{eq:3}
\end{eqnarray}
Here $\rd V_{\Gamma}$ denotes the volume element on $\Gamma$. In a
microcanonical equilibrium the temperature is determined
intrinsically by the thermodynamic relation $T\rd S=\rd E$, which
implies that $k_B T =\Omega(E)/\Omega^\prime(E)$, where
$\Omega^\prime(E)=\rd\Omega(E)/\rd E$. Since the density of states
$\Omega(E)$ is differentiable, the temperature is well-defined.
Other thermodynamic quantities can likewise be precisely determined.
For example, the specific heat $C(T)=\rd E/\rd T$ is given by $C=
k_B (\Omega^\prime)^2/[(\Omega^\prime)^2 - \Omega
\Omega^{\prime\prime}]$.

Consider a large system composed of two independent parts, each in
a state of equilibrium. Each subsystem is thus described by a
microcanonical state with support on the Segr\'e variety
corresponding to disentangled subsystem states. Let us write
$\Omega_1(E_1)$ and $\Omega_2(E_2)$ for the associated state
densities, where $E_1$ and $E_2$ are the initial energies of the
two systems. Now imagine that the two systems interact weakly for
a period of time, during which energy is exchanged, following
which the systems become independent again, each in a state of
equilibrium. As a consequence of the interaction the state
densities of the systems will now be given by expressions of the
form $\Omega_1(E_1 + \epsilon)$ and $\Omega_2(E_2-  \epsilon)$,
for some value of the exchanged energy  $\epsilon$. The value of
$\epsilon$ can be determined by the requirement that the total
entropy $S(E)=k_B\ln[\Omega_1(E_1 + \epsilon)\Omega_2(E_2-
\epsilon)]$ should be maximised. A short calculation shows that
this condition is satisfied if and only if $\epsilon$ is such that
the temperatures of the two systems are equal. This argument shows
that the definition of temperature that we have chosen is a
natural one, and is physically consistent with the principles of
equilibrium thermodynamics.

\textbf{Phase transitions}. The quantum microcanonical ensemble
introduced here is applicable to any isolated finite-dimensional
quantum system for which the ideal gas approximation is valid. The
volume integral in (\ref{eq:3}) can be calculated by lifting the
integration from $\Gamma$ to ${\mathcal H}$ and imposing the
constraint that the norm of $|\psi\rangle$ is unity. Then we can
write:
\begin{eqnarray}\label{eq:int2}
\Omega(E)=\frac{1}{\pi}\int_{{\mathcal H}}\! \delta( \langle \psi
|\psi\rangle -1)\;\delta\Big(\frac{\langle \psi | {\hat
H}|\psi\rangle}{\langle \psi |\psi\rangle}-E\Big) {\rm d}
V_{\mathcal H}, \label{eq:3-1}
\end{eqnarray}
where ${\rm d} V_{\mathcal H}$ is the volume element of ${\mathcal
H}$. Making use of the standard Fourier integral representation
for the delta function, and diagonalising the Hamiltonian, we find
that (\ref{eq:3-1}) reduces to a series of Gaussian integrals (see
\cite{brody0} for details). Performing the $\psi$-integration we
then obtain the following integral representation for the density
of states:
\begin{eqnarray}
\Omega(E)=(-\ri\pi)^n\!\!\int\limits_{-\infty}^\infty\!
\frac{\rd\nu}{2\pi} \int\limits_{-\infty}^\infty\!
\frac{{\rd}\lambda}{2\pi\ri} \, \re^{\ri(\lambda+\nu E)}
\prod_{l=1}^{n+1}\frac{1} {(\lambda+\nu E_l)}. \label{eq:6}
\end{eqnarray}
The two integrals appearing here correspond to the delta-functions
associated with the energy constraint $H(\psi)=E$ and the norm
constraint $ \langle \psi|\psi\rangle =1$. Carrying out the
integration we find that the density of states is given by
\begin{eqnarray}
\Omega(E)&=&\frac{(-1)^{m-1}\pi^{n}}{(n-1)!} \prod_{j=1}^{m}
\frac{1}{(\delta_j-1)!} \left( \frac{\rd}{\rd E_j}
\right)^{\delta_j-1} \nonumber \\ && \times \sum_{k=1}^{m}
(E_k-E)^{n-1} \prod_{l\neq k}^{m} \frac{{\mathbf 1}_{\{E_k>
E\}}}{E_l-E_k}, \label{eq:7}
\end{eqnarray}
where ${\mathbf 1}_{\{A\}}$ denotes the indicator function
(${\mathbf 1}_{\{A\}}=1$ if $A$ is true, and $0$ otherwise). In
(\ref{eq:7}) we let $m$ denote the number of distinct eigenvalues
$E_j$ ($j=1,2,\ldots,m$), and we let $\delta_j$ denote the
multiplicity associated with the energy $E_j$. Thus $\sum_{j=1}^m
\delta_j=n+1$. In the nondegenerate case, for which $\delta_j=1$
for $j=1,2,\ldots,m$, we have
\begin{eqnarray}
\Omega(E)=\frac{(-\pi)^n}{(n-1)!} \sum_{k=1}^{n+1} (E_k-E)^{n-1}
\prod_{l\neq k}^{n+1} \frac{{\mathbf 1}_{\{E_k> E\}}}{E_l-E_k}.
\label{eq:8}
\end{eqnarray}
With these expressions at hand we proceed now to examine some
explicit examples.

\begin{figure}[th]
{\centerline{\psfig{file=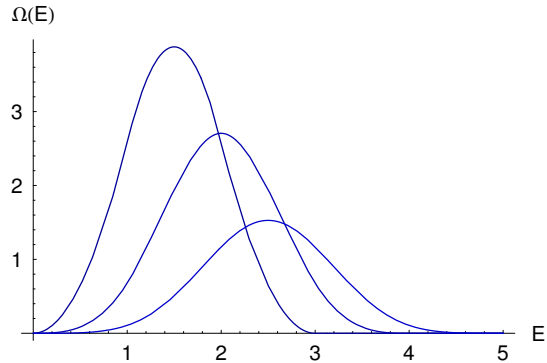,width=8cm,angle=0}}}
 \caption{Density of states $\Omega(E)$ for nondegenerate
 $(n+1)$-level systems for $n=3,4,5$, with energy eigenvalues
 $E_k=k$ $(k=0,\ldots,n)$. $\Omega(E)$ is $n-2$ times
 differentiable in $E$.
  \label{fig:1}}
\end{figure}

\textbf{Nondegenerate spectra}. In the case of a Hamiltonian with
a nondegenerate spectrum of the form $E_k= \varepsilon(k-1)$,
$k=1,2,\ldots, n+1$, where $\varepsilon$ is a fixed unit of
energy, the density of states (\ref{eq:8}) reduces to
\begin{eqnarray}
\Omega(E) = \varepsilon^{-1}\frac{(-\pi)^n}{(n-1)!}\!\sum_{k >
E/\varepsilon}^{n} \frac{(-1)^k \left(k-E/\varepsilon
\right)^{n-1}}{ k!(n-k)!}. \label{eq:9}
\end{eqnarray}
We see that $\Omega(E)$ is a polynomial of degree $n-1$ in each
interval $E\in[E_j,E_{j+1}]$, and that for all values of $E$ it is
at least $n-2$ times differentiable. In Fig.~\ref{fig:1} we plot
$\Omega(E)$ for several values of $n$. For a system in equilibrium
the accessible values of $E$ are those for which $\Omega^\prime(E)
\geq0$. States for which $\Omega^\prime(E)<0$ have ``negative
temperature'' in the sense of Ramsey~\cite{ramsey}.

The structure of the space of pure states in quantum mechanics is
intricate, even for relatively elementary systems. In particular,
as the value of the energy changes, the topological structure of
the energy surface undergoes a transition at each
eigenvalue~\cite{brody1}. For example, in the case of a
nondegenerate three-level system, the topology of the energy
surface changes according to: ${\rm Point}\to S^3\to S^1\times
{\mathbb R}_{\#}^2 \to S^3 \to {\rm Point}$, as the energy is
raised from $E_{\rm min}$ to $E_{\rm max}$ (${\mathbb R}_{\#}^2$
denotes a two-plane compactified into $S^2$ at a point
corresponding to the intermediate eigenstate). These structural
changes in the energy surfaces induce a corresponding nontrivial
behaviour in the thermodynamic functions.

\begin{figure}[th]
{\centerline{\psfig{file=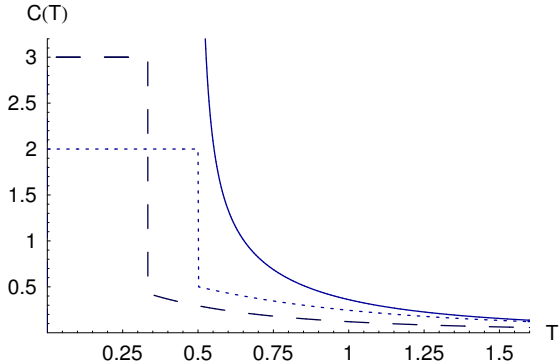,width=8cm,angle=0}}}
 \caption{Specific heat for a nondegenerate four-level system
 (dotted line, $n=3$, $E_j=0,1,2,3$), a four-level system
 having a degenerate first excited state (dashed line, $n=4$,
 $E_j=0,1,1,2,3$), and a quantum Ising chain (solid line, $J=1/4$,
 $B=1$). In the quantum Ising chain, we have $C(T)\sim
 (T-T_c)^{-2}$ away from $T_c$, whereas in the vicinity of $T_c$ we
 have $C(T)\sim (T-T_c)^{-13}$ for $T>T_c$. (We set
 $k_B=1$ here.)
 \label{fig:2}}
\end{figure}

As an illustration we consider a four-level system and compute the
specific heat as a function of temperature. The result is shown in
Fig.~\ref{fig:2}, where we observe that the specific heat drops
abruptly from $2k_B$ to $\frac{1}{2}k_B$ at the critical
temperature $T_c$ defined by $k_BT_c=\frac{1}{2}\varepsilon$.
Therefore, this system exhibits a second-order phase transition,
in this case at the critical energy $E_c=\varepsilon$. This
example shows that the relationships between phase transitions and
topology discovered recently in classical statistical mechanics
\cite{franzosi} carry over to the quantum domain where, arguably,
they may play an even more basic role.

For a system with a larger number of nondegenerate eigenstates,
the specific heat also increases abruptly as $T$ is reduced. In
this case the specific heat is continuous, and the discontinuity
is in a higher-order derivative of the energy. For a system with
$n+1$ nondegenerate energy eigenvalues, the $(n-1)$-th derivative
of the energy with respect to the temperature has a discontinuity.
The phenomenon of a continuous phase transition is generic, and is
also observed if the eigenvalue spacing is not uniform.

\textbf{Degenerate spectra}. In a system with a degenerate
spectrum, the phase transition can be enhanced. In particular, the
volume of ${\mathcal E}_E$ increases more rapidly as $E$
approaches the first energy level from below, if this level is
degenerate. This leads to a more abrupt drop in the specific heat
(Fig.~\ref{fig:2}).

As an example, we consider a quantum Lenz-Ising ferromagnetic
chain with three spins. The Hamiltonian is ${\hat H} = -J
\sum_{k=1}^3 \sigma_z^k \sigma_z^{k+1} - B \sum_{k=1}^3
\sigma_z^k$, where $\sigma_z^k$ denotes the third Pauli matrix for
spin $k$, and $J,B$ are constants. We have in mind a gas of weakly
interacting molecules, each modelled by a strongly-interacting
quantum Ising chain. The eigenvalues of the Hamiltonian are
$E_1=-3J-3B$, $E_{2,3,4}=J-B$, $E_{5,6,7}=J+B$, and $E_8=-3J+3B$.
As the temperature is reduced, the specific heat grows rapidly in
the vicinity of the critical point $T_c=(2J+B)/3k_B$
(Fig.~\ref{fig:2}), where the system exhibits a discontinuity in
the second derivative of the specific heat. We note that when $B$
is small the critical temperature is close to that of the
classical mean-field Ising model.

\textbf{Density matrix and energy uncertainty}. Finally, we show the
existence of a natural energy band associated with the quantum
microcanonical distribution. The microcanonical density matrix for
the energy $E$ is
\begin{eqnarray}
{\hat\mu}_E = \frac{1}{\Omega(E)} \int_{\Gamma} \delta(H(\psi)-E)
{\hat \Pi}(\psi)\, \rd V_{\Gamma}. \label{eq:12}
\end{eqnarray}
Here ${\hat \Pi}(\psi)=|\psi\rangle\langle\psi | /\langle\psi|
\psi\rangle$ denotes the projection operator onto the state
$|\psi\rangle\in{\mathcal H}$ corresponding to the point
$\psi\in\Gamma$. The squared energy uncertainty is $(\Delta
H)^2={\rm tr} ({\hat\mu}_E{\hat H}^2)-[{\rm tr}({\hat\mu}_E {\hat
H})]^2$.  A  calculation then shows that
\begin{eqnarray}
(\Delta H)^2= \frac{n+1}{\Omega(E)} \int_{E_{\rm min}}^{E}\!({\bar
H} -u)\, \Omega(u)\, \rd u, \label{eq:13}
\end{eqnarray}
where ${\bar H}={\rm tr}({\hat H})/(n+1)$ denotes the uniform
average of the energy eigenvalues. To check that $\Delta H$
vanishes at $E=E_{\rm max}$ we note that the first moment of
$\Omega(E)$ is given by the integral of $H(x)$ over $\Gamma$.
Hence by use of a trace identity obtained in \cite{gibbons} we
have
\begin{eqnarray}
\int_{E_{\rm min}}^{E_{\rm max}}\!u\,\Omega(u)\,\rd
u=\int_{\Gamma} H(\psi)\,\rd V_{\Gamma} = \frac{\pi^n}{n!}\,{\bar
H}. \label{eq:14}
\end{eqnarray}
However, the integral of $\Omega(E)$ is the volume $\pi^n/n!$ of
$\Gamma$, and the desired result follows. Using the explicit
formulae obtained earlier for $\Omega(E)$ we are then able to
calculate the energy uncertainty associated with the equilibrium
state of a finite quantum system. It remains to be seen whether the
new ensemble can be put to the test in some definitive way, and in
particular whether the phase transitions it predicts actually
correspond to observable phenomena.

DCB acknowledges support from The Royal Society. The authors thank
M.~Kastner, T.W.B.~Kibble, and an anonymous referee for helpful
comments.

\end{document}